\documentclass[a4paper,10pt,twoside]{cpc-hepnp}
\usepackage{CJK,upgreek,fancyhdr}
\usepackage{multicol}
\usepackage{graphicx}
\usepackage{booktabs}
\usepackage{amssymb,bm,mathrsfs,bbm,amscd}
\usepackage[tbtags]{amsmath}
\usepackage{lastpage}

\begin{document}
%\begin{CJK*}{GB}{gbsn}
%\begin{CJK*}{GBK}{song}

\fancyhead[c]{\small Submitted to Chinese Physics C}
%\fancyfoot[C]{\small 010201-\thepage}

%\footnotetext[0]{Received 31 June 2015}

\title{Preliminary test results of LAr prototype detector}

\author{%
Li Pei-Xian$^{1,2}$\
\quad Guan Meng-Yun$^1$
\quad Yang Chang-Gen$^1$
\quad Zhang Peng$^1$\\
\quad Liu Jin-Chang$^1$
\quad Zhang Yong-Peng$^{1,2}$
\quad Guo Cong$^{1,2}$
\quad Wang Yi$^{1,2}$
}
\maketitle

\address{%
$^1$Institute of High Energy Physics, CAS, Beijing 100049, China\\
$^2$Graduate University of Chinese Academy of Sciences, Beijing 100049, China\\
}

\begin{abstract}
WIMPs are a well-motivated galactic dark matter candidate. Liquid argon (LAr) is an attractive target for the direct detection of WIMPs. The LAr prototype detector is designed to study the technology and property of LAr detector. The prototype detector have an active volume containing 0.65 kg of liquid argon. The liquid nitrogen(LN) cooling system allows the temperature of liquid argon to be maintained at the boiling point (87.8 K) with fluctuations less than 0.1 K. The prototype was calibrated with a Na$^{22}$ source, with the light yield 1.591$\pm$0.019 p.e./keV for the 511 keV gamma rays using the domestic-made argon purification system.
\end{abstract}

\begin{keyword}
LAr technology, scintillation light, light yield, Dark Matter
\end{keyword}

\begin{pacs}
95.35.+d, 29.40.-n
\end{pacs}

%\footnotetext[0]{\hspace*{-3mm}\raisebox{0.3ex}{$\scriptstyle\copyright$}2013
%Chinese Physical Society and the Institute of High Energy Physics
%of the Chinese Academy of Sciences and the Institute
%of Modern Physics of the Chinese Academy of Sciences and IOP Publishing Ltd}%

\begin{multicols}{2}

\section{Introduction}
  The data of Planck \cite{Plank} provide compelling evidence for a significant cold dark matter component in the composition of the Universe.Weakly interacting massive particles(WIMPs) are a well-motivated galactic dark matter candidate. Numerous direct detection experiments\cite{experiment1,experiment2,experiment3,experiment4,experiment5} are being developed to detect WIMPs. Liquid argon (LAr) detectors are attractive detectors for the direct detection of WIMPs\cite{Lar2}.Liquid argon is excellent scinilation materials, with a high light yield of approximately 40 photons per keV\cite{LY}.

  Liquid argon provides outstanding pulse-shape discrimination(PSD) based on scintillation timing. The excitation and ionization of the medium from particles interacting with argon atom will lead to two excited state. The two excited states have different lifetimes, about 6 ns for the singlet state and 1.6 $\mu$s for the triplet state\cite{lifetimes}. With sufficient photon statistics, PSD allows discrimination of nuclear recoil events from electron-induced background events at better than $10^{8}$\cite{experiment4,PSD2,PSD3} which provide a way to detect rare nuclear recoil events.

\section{The LAr prototype detector design and structure}
  The liquid Argon prototype detector, shown in Fig. \ref{fig:LAR}, consists of a double wall vacuum stainless steel cryostat and an inner vessel which contains 0.65 kg active liquid argon. This inner vessel is completely immersed in a liquid argon bath contained in the cryostat. The inner vessel consists of a PTFE cylinder and two fused silica windows. The cylinder is 9.2 cm height, 8.0 cm inner diameter, and 1.0 cm wall thickness. The PMTs are in the outer LAr bath, viewing the active volume through the top and bottom fused silica windows which are the top and bottom lids of the inner vessel.

  The center wave length of the argon scintillation light is 128 nm, so the 1,1,4,4-Tetraphenyl-1,3-butadiene (TPB) is needed as the wavelength shifter to convert the 128 nm scintillation photons to the wavelength for detection by photomultipliers. The peak of emission wavelength of the TPB is 420 nm\cite{TPB}. The TPB is deposited onto the inner surface of the PTFE acrylic cylinder and the inner surfaces of the fused silica windows in a thermal vacuum coating device. The PTFE cylinder and fused silica windows were coated with about 200 $\mu$g cm$^{2}$ of TPB. During the evaporation we took care to protect the outer surface of the PTFE cylinder and fused silica windows from the TPB steam, because the LAr scintillation from the inactive region will be transfer to the wavelength-shifted light by the TPB attached to the outer surface of the PTFE cylinder and fused silica windows, finally the light could be collected by the PMTs. After the evaporation the parts were kept in sealed bags filled with dry argon.

   \begin{center}
  \includegraphics[width=8cm]{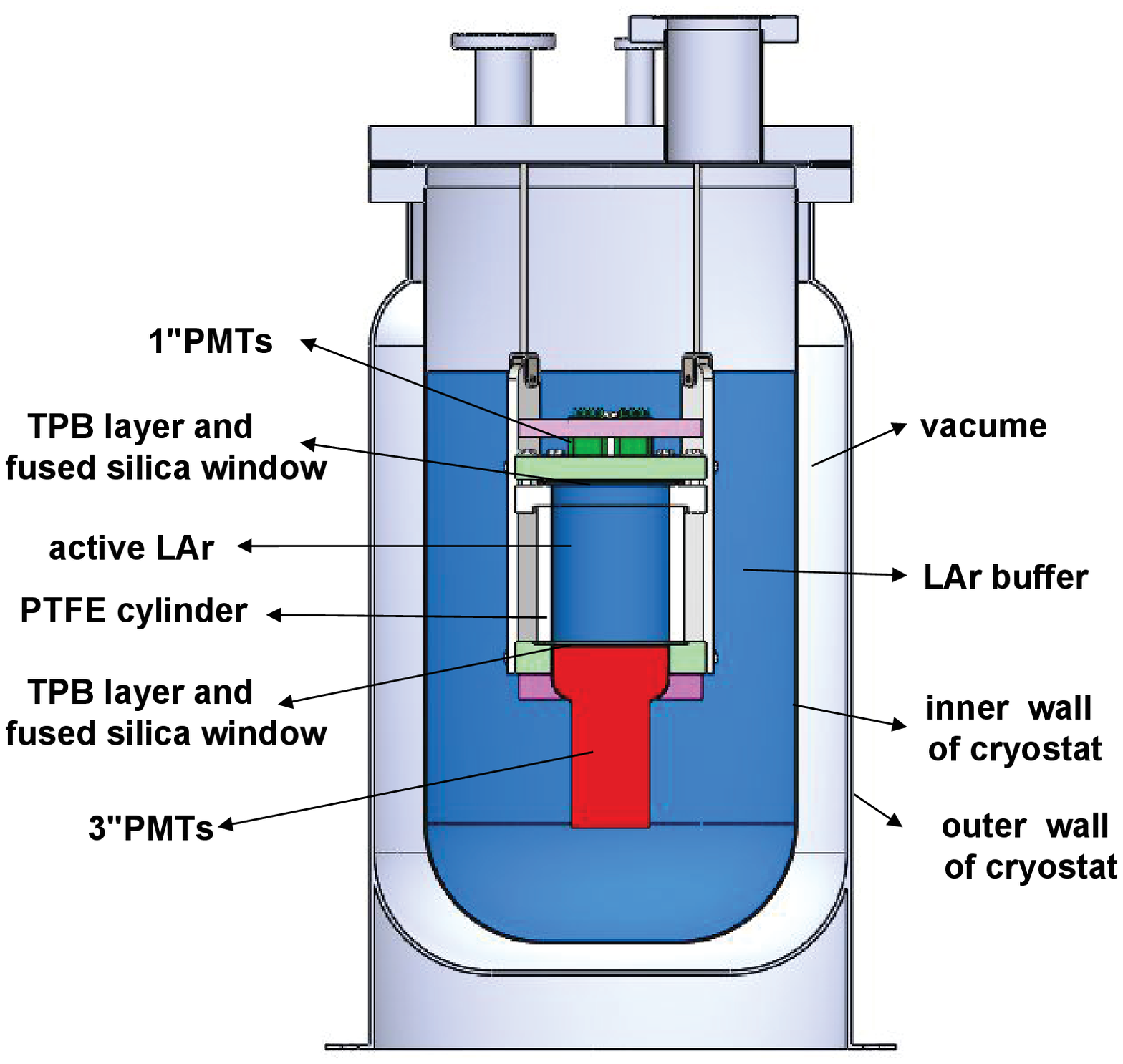}
  \figcaption{\label{fig:LAR} The liquid Argon prototype detector structure. The outer layer is the double vacuum stainless steel cryostat.The inner vessel is immersed in a liquid argon bath. The rad and green parts are the 3" PMT and 1" PMTs. The blue parts are the liquid argon.
  }
  \end{center}

  The wavelength-shifted  light is collected by two arrays of PMTs, shown in Fig. \ref{fig:LAR}, viewing the active volume through the top and bottom fused silica windows. Three HAMAMATSU R8520-06mod PMTs(1 inch) are in the top array, and one HAMAMATSU R11065 PMT(3 inch) is in the bottom array. To reduce escape of the wavelength-shifted light from the inner vessel, the spaces between the PMTs are filled with PTFE reflectors. About 1 mm layer of LAr optically couples the PMTs to the windows. The room-temperature quantum efficiencies of the R11065 is 25$\%$ at 420 nm\cite{PMT}. They are run at a typical gain of 4.2$\times$10$^{6}$.
  \section{Cooling, purification and recirculation system}

  The liquid nitrogen(LN) cooling system, shown in Fig.\ref{fig:lab}, is consist of a 100L liquid nitrogen Dewar and a cold-head. The boiling point of liquid nitrogen is about 10K at the same pressure lower than the boiling point of the liquid argon, so we can use the liquid nitrogen to liquefy the gas argon. The cold-head is inside the cryostat but outside the inner vessel. The 100L liquid nitrogen Dewar is connected to a cold-head through a lengthy bellow. The liquid nitrogen go in to the cold head left the latent heat to give the cooling capacity. We can adjust the cooling power though controlling the vaporized nitrogen flow using the pressure of the LAr cryostat as the process variable,We can use a proportional每integral每derivative(PID) controller to give control to the temperature of the LAr. Test result show the LAr temperature fluctuation is less than 0.1K.
     \begin{center}
  \includegraphics[width=8cm]{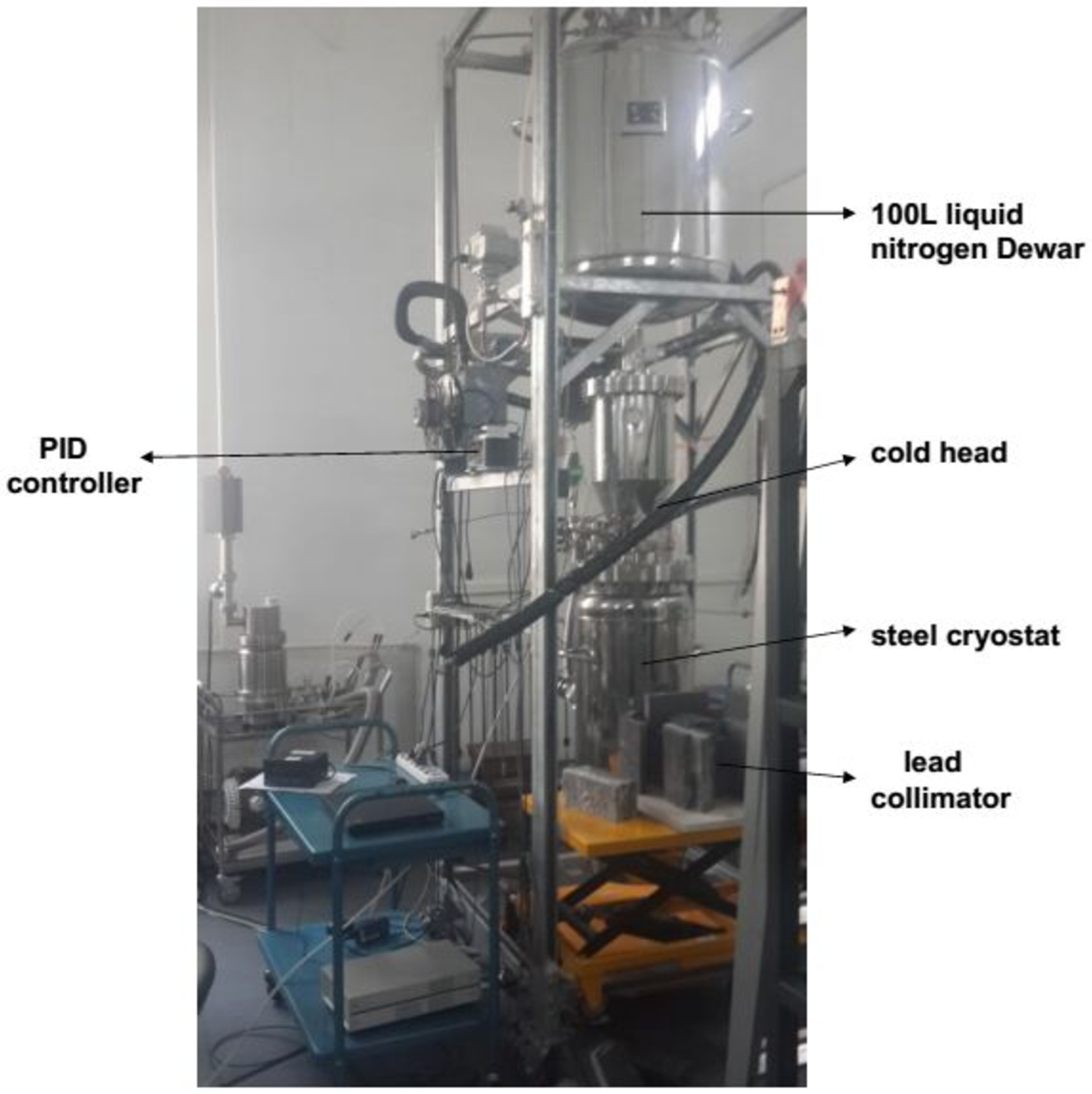}
  \figcaption{\label{fig:lab} The LN cooling system together with the prototype detector system.
  }
  \end{center}

  To  obtain the maximum light yield,the dissolved electronegative impurities such as nitrogen, oxygen, and water need to be lower than 0.1 ppm\cite{scintillation}. A special domestic-make purification system, which is developed by Beijing Beiyang United Gas Co.,Ltd, combined with a gas recirculation loop to fulfilled the demand. Before filling the argon, the LAr cryostat is pumped for several days, till the pressure of the cryostat maintained at about 10$^{-4}$ pa. Then the feed argon gas (95$\%$) is purified by a purification system and flow to the LAr cryostat, shown in Fig.\ref{fig:getter}. After filling the liquid argon, we start the argon recirculation to further improve the argon purity.Liquid argon is vaporized by a heater installed at the bottom of the LAr cryostat. a senior aerospace metal bellows pump is used to drive the argon gas to recirculate through the purification system, to flow back to the cold-head to be liquified again. Argon gas passed through the purification system several times until the scintillation properties were not improved further.
       \begin{center}
  \includegraphics[width=8cm]{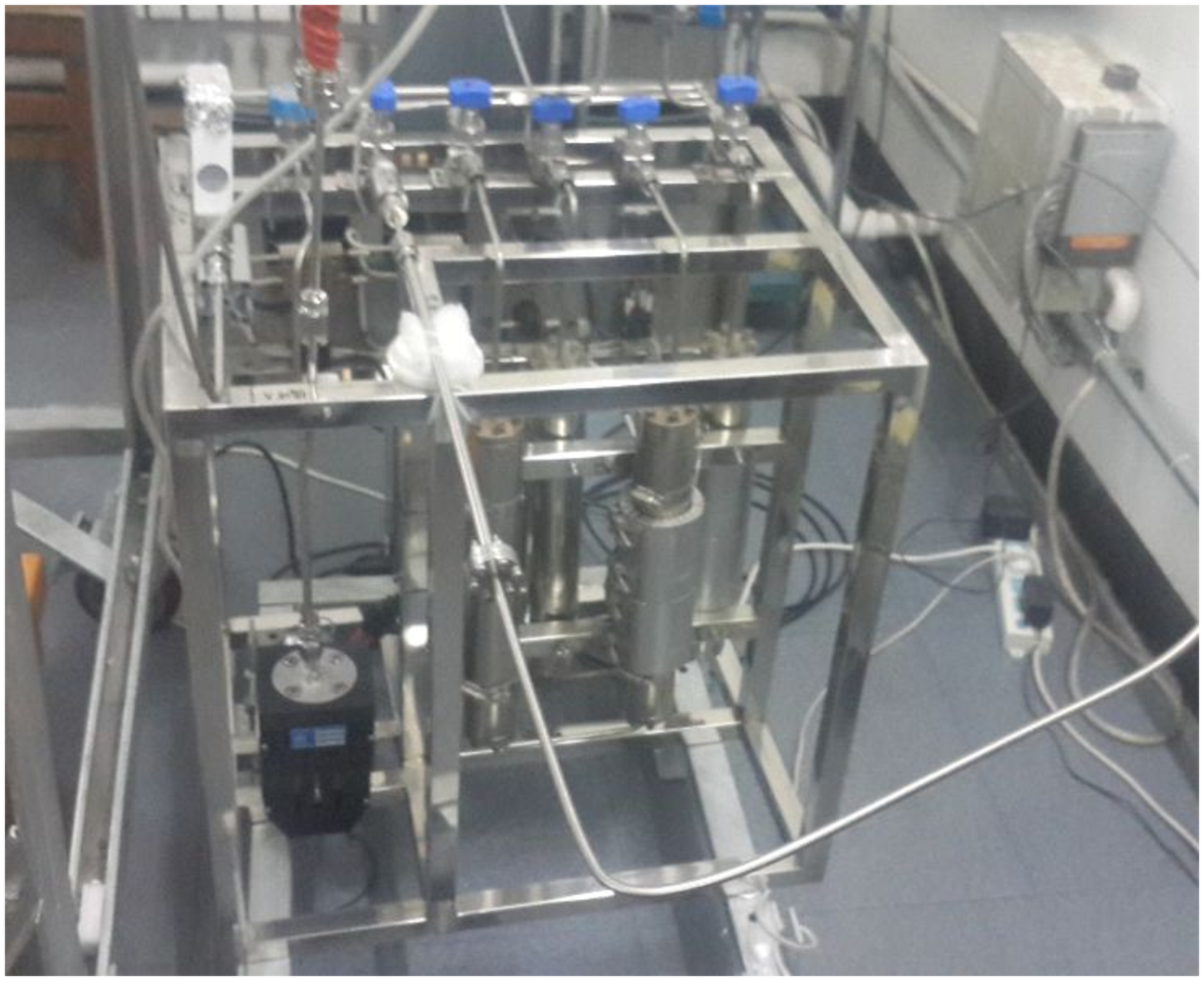}
  \figcaption{\label{fig:getter} The getter developed by Beijing Beiyang United Gas Co.,Ltd is designed for reducing nitrogen, oxygen, and water impurities to lower than 0.1 ppm levels.
  }
  \end{center}

  \section{DAQ system and single-photoelectron calibration}
  We use LeCroy WAVERUNNER 610Zi oscilloscope\cite{DAQ} to record the signals waveform from the 3" and 1§ photomultiplier tubes for off-line analysis. The trigger required a coincidence of one 1" PMT and one 3" PMT signals in 300 ns. The threshold of the 3§ PMT is 10mV, and that of the 1§ PMT is 6mV. When an event satisfies the trigger condition, data in a 10 $\mu$s time window (1.6 $\mu$s before the trigger, 8.4 $\mu$s after the trigger gate), is stored on a local hard disk.

  We use LED installed in the cryostat to do the single-photoelectron calibration of the 3" PMT. The pulse width for LED run is 200 ns. The trigger is generated by a pulse generator. The spectra and the fit of the single- photoelectron is shown in Fig.\ref{fig:getter}. we obtained about1.436$\pm$0.021 pC/PE.
       \begin{center}
  \includegraphics[width=8cm]{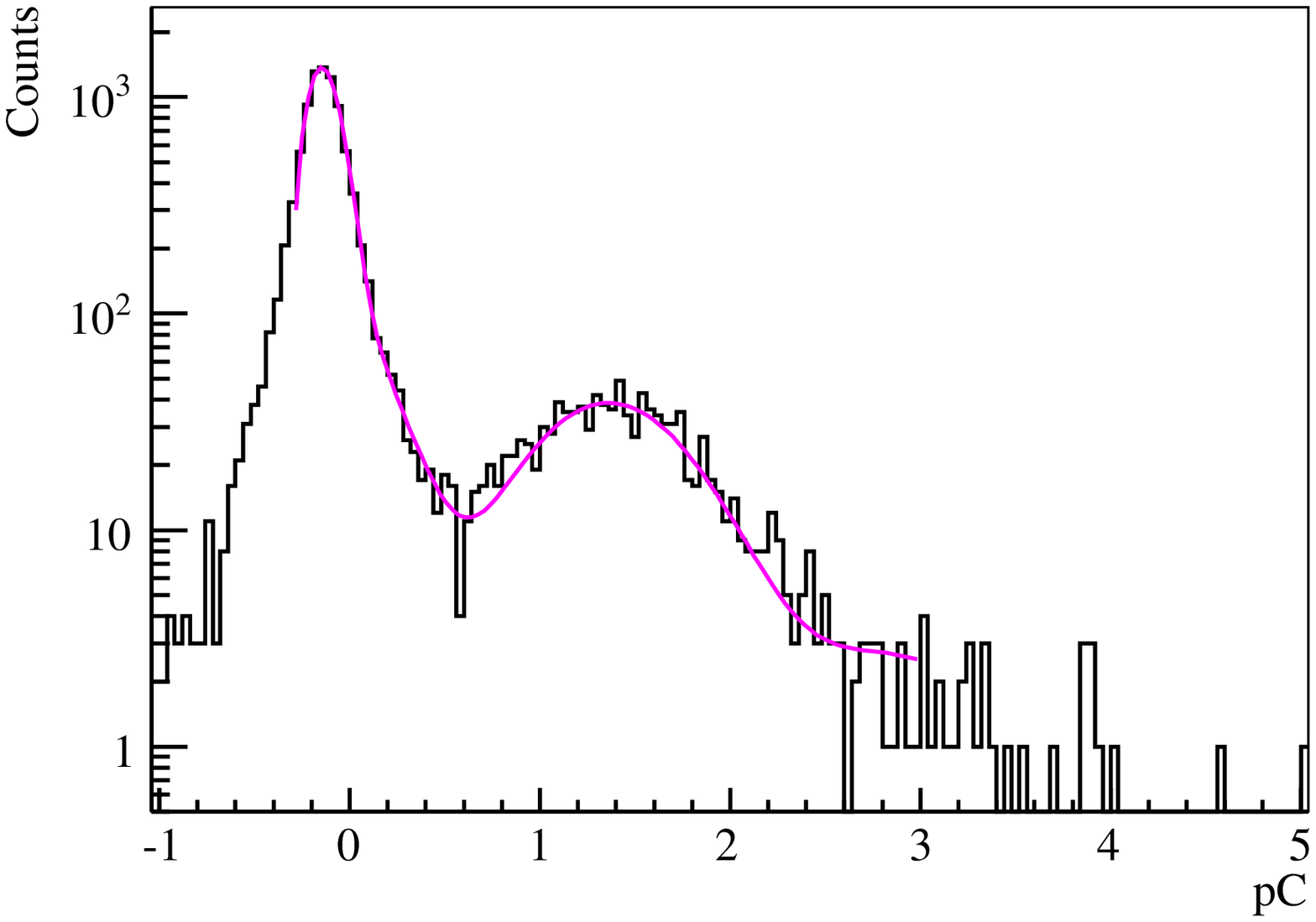}
  \figcaption{\label{fig:spe} Single photoelectron spectra of the 3" PMT, used to calibrate the absolute light yield and number of detected photons..
  }
  \end{center}

\section{Event analysis and light yield}

 We determine a baseline and subtract it from the waveform for the two individual channels. The average of the digitized samples in the 1.6$\mu$s pro-trigger (where no signal is expected) was calculated. Once the baseline has been subtracted, the integral is evaluated for the waveform to give the charge of the event.

  The detector is exposed to a 70$\mu$Ci  $^{22}$Na gamma source. The $^{22}$Na source emit two 510.99 keV gammas with opposite directions. So, we can use this two gammas as a coincidence. The 70$\mu$Ci $^{22}$Na gamma source is collimated by a thick lead collimator which aim at the center of the inner vessel. On the other side of the $^{22}$Na source, another collimator is aimed at a plastic scintillator, which is used as the coincidence detector. The plastic scintillator coat with the black tape as the reflector and use the one 3" PMT as the read out.

    \begin{center}
  \includegraphics[width=8cm]{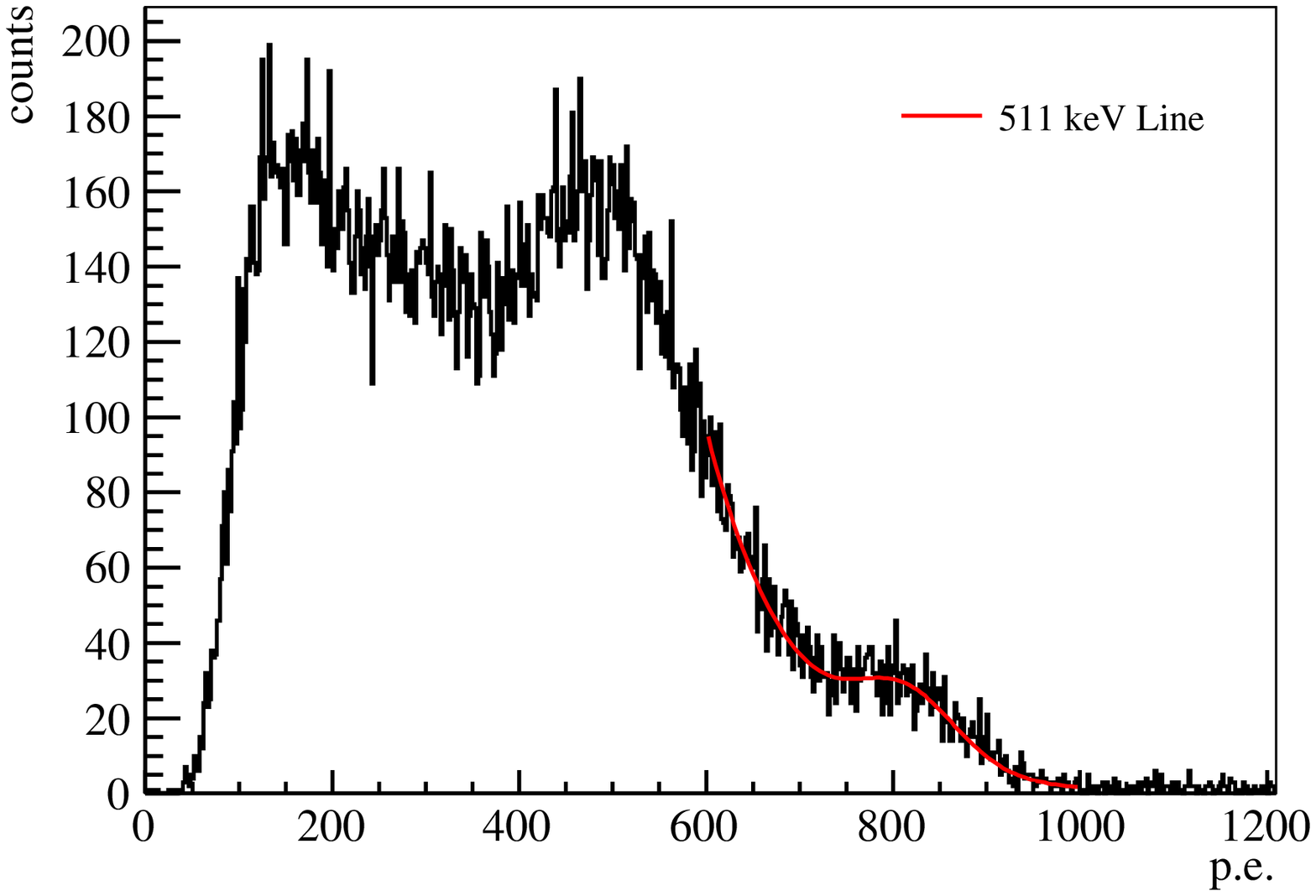}
  \figcaption{\label{fig:Na22_0.511MeV}   Scintillation spectrum of $^{22}$Na coincidence with a plastic scintillator detector. Red: the 0.511MeV full-energy peak fitted with a Gaussian plus a falling exponential function.
   }
  \end{center}

   Fig. \ref{fig:Na22_0.511MeV} shows the gamma-induced scintillation spectra of 50000 events. Due to the small size of the active volume(0.65kg), the 511 keV $\gamma$ is difficult to deposit its full energy in the detector. So the spectra is degraded leaving the Compton edge much more visible than the full-energy peak. This agree with the expectations from a GEANT4-based Monte Carlo simulation of the experimental setup, including the material between the source and the active volume. The full-energy peak is fit with the sum of a Gaussian and a falling exponential. The energy resolution of 511 keV $\gamma$ rays is 6.75 $\%$.

          \begin{center}
\tabcaption{ \label{tab1:LY} The peak mean, width, and light yield of the gamma full-absorption peak. The error on $\mu$ is the statistical error from the fit. The error on LY is the fit error combined with the statistical error on the mean single-p.e. response.}
\footnotesize
\begin{tabular*}{80mm}{c@{\extracolsep{\fill}}ccccc}
\toprule
    E$_{\gamma}$[keV]           &$\mu$  &$\sigma$  &LY \\
    \hline
    510.99  &813.21$\pm$3.79   &54.90  &1.591$\pm$0.019   \\
\bottomrule
\end{tabular*}
\end{center}

\section{Conclusion}

 Using the customized LN cooling system,and the domestic made argon purification system, we set up a stable LAr prototype detector. With a collimated $^{22}$Na source, a proper energy spectra showed about  1.591$\pm$0.019 p.e./keV light yield for 0.511 MeV gammas.

\end{multicols}
\vspace{-1mm}
\centerline{\rule{80mm}{0.1pt}}
\vspace{2mm}

\begin{multicols}{2}

\end{multicols}

\clearpage
%\end{CJK*}
\end{document}